\documentstyle[12pt]{article}
\begin{document}

\title{\Large\bf Dynamical Correlations for Circular Ensembles \\ 
of Random Matrices}
\author{Taro Nagao and Peter J. Forrester$^{\dagger}$}
\date{}
\maketitle

\begin{center}
\it
Department of Physics, Graduate School of Science, Osaka University, \\
Toyonaka, Osaka 560-0043, Japan
\end{center}
\begin{center}
\it
 $^{\dagger}$Department of Mathematics and Statistics, University of Melbourne, \\
Victoria 3010, Australia
\end{center}

\bigskip
\begin{center}
\bf Abstract 
\end{center}
\par
\bigskip
\noindent
Circular Brownian motion models of random matrices were 
introduced by Dyson and describe the parametric eigenparameter 
correlations of unitary random matrices. For symmetric unitary, 
self-dual quaternion unitary and an analogue of antisymmetric hermitian 
matrix initial conditions, Brownian dynamics toward the 
unitary symmetry is analyzed. The dynamical correlation 
functions of arbitrary number of Brownian particles at 
arbitrary number of times are shown to be written in 
the forms of quaternion determinants, similarly 
as in the case of hermitian random matrix models. 

\par
\bigskip
\bigskip
\noindent
{\it PACS}: 05.20.-y; 05.40.-a; 02.50.Ey
\par
\bigskip
\noindent
{\it KEYWORDS}: random matrix; Fokker-Planck equation; 
quaternion determinant

\newpage
\noindent
\section{Introduction}
\setcounter{equation}{0}
\renewcommand{\theequation}{1.\arabic{equation}}
Circular ensembles of random matrices were introduced 
by Dyson as the simplest possible models of complex 
energy spectra\cite{DYC,MEHTA1}. Dyson further introduced a Brownian 
motion model to describe the parametric correlation of 
the energy levels\cite{DYB}. His Brownian dynamics is specified 
by the Fokker-Planck equation
\begin{equation}
\frac{\partial p}{\partial \tau} = {\cal L} p, \ \ \ 
{\cal L} = \sum_{j=1}^N \frac{\partial}{\partial \theta_j} \left( 
\frac{\partial W}{\partial \theta_j} + \frac{1}{\beta} 
\frac{\partial}{\partial \theta_j} \right),
\label{FOKKER}
\end{equation} 
where $\theta_1,\theta_2,\cdots,\theta_N$ are the locations of Brownian 
particles, $\tau$ is the time variable and 
\begin{equation}
W = - \sum_{j<l}^N \log\left| {\rm e}^{\theta_j} - {\rm e}^{\theta_l} \right|.
\end{equation}
The stationary distribution of this Brownian dynamics is 
\begin{equation}
p_S = {\rm e}^{-\beta W} = \prod_{j<l}^N 
\left| {\rm e}^{\theta_j} - {\rm e}^{\theta_l} \right|^{\beta}
\end{equation}
satisfying $\partial p_S/\partial \tau = 0$. Using a perturbation 
theory, Dyson showed that it 
is a natural model to describe the parametric eigenparameter 
correlation of symmetric unitary, unitary and self-dual quaternion 
unitary random matrices, corresponding to $\beta=1,2$ and $4$, 
respectively. 
\par
In order to solve the Fokker-Planck equation, a transformation of 
the Fokker-Planck operator ${\cal L}$ into a hermitian operator ${\cal H}$ 
is useful. The hermitian operator ${\cal H}$ is defined as
\begin{equation}
{\rm e}^{\beta W/2} {\cal L} {\rm e}^{-\beta W/2} = - \frac{1}{\beta} 
({\cal H} - E_0)
\label{LANDH}
\end{equation}
with a constant $E_0$ and explicitly written as
\begin{equation}
{\cal H} = - \sum_{j=1}^N \frac{\partial^2}{\partial \theta_j^2} + 
\frac{\beta (\beta/2 - 1)}{4}  \sum_{j<l}^N \frac{1}{
\sin^2[(\theta_j - \theta_l)/2]}.
\label{CS}
\end{equation}
It is known as the Calogero-Sutherland Hamiltonian on the unit circle. 
This Hamiltonian has a complete set of orthogonal eigenfunctions 
\begin{equation}
\psi_{\kappa}(\theta_1,\cdots,\theta_N) 
= {\rm e}^{- \beta W/2} P^{(2/\beta)}_{\kappa}(z_1,\cdots,z_N), \ \ \ z_j 
= {\rm e}^{i \theta_j}
\end{equation}
with eigenvalues $E_{\kappa}$\cite{SUTH,PJF1,FNPOI}. 
Here $\kappa = (\kappa_1,\cdots,\kappa_N)$ represents a partition with 
non-negative integers $\kappa_1 \geq \kappa_2 \geq 
\cdots \geq \kappa_N$ and $P^{(2/\beta)}_{\kappa}(z_1,\cdots,z_N)$ 
is a particular polynomial known as the Jack polynomial.
\par
For the imaginary time Schr\"odinger equation
\begin{equation}
\frac{\partial \psi}{\partial \tau} = - \frac{1}{\beta} (H - E_0) \psi,
\end{equation}
the Green function solution can be written as
\begin{equation}
G^{(H)}(\varphi_1,\cdots,\varphi_N;\theta_1,\cdots,\theta_N;\tau) = 
\sum_{\kappa} \frac{{\bar \psi}_{\kappa}(\varphi_1,\cdots,\varphi_N) 
\psi_{\kappa}(\theta_1,\cdots,\theta_N)}{\langle \psi_{\kappa} | 
\psi_{\kappa} \rangle} {\rm e}^{-(E_{\kappa} - E_0)\tau/\beta},
\end{equation}
where
\begin{equation}
\langle \psi_{\kappa} | \psi_{\kappa} \rangle = 
\int_{-\pi}^{\pi} {\rm d}\theta_1 \cdots 
\int_{-\pi}^{\pi} {\rm d}\theta_N |\psi_{\kappa}(\theta_1,\cdots,\theta_N)|^2
\end{equation} 
and ${\bar \psi}_{\kappa}$ is the complex conjugate of $\psi_{\kappa}$. 
According to (\ref{LANDH}), the Green function solution 
$G(\varphi_1,\cdots,\varphi_N;\theta_1,\cdots,\theta_N;\tau)$ of 
the Fokker-Planck equation (\ref{FOKKER}) is given by 
\begin{equation}
G(\varphi_1,\cdots,\varphi_N;\theta_1,\cdots,\theta_N;\tau) = 
\frac{{\rm e}^{- \beta W(\theta_1,\cdots,
\theta_N)/2}}{{\rm e}^{- \beta W(\varphi_1,\cdots,\varphi_N)/2}}
G^{(H)}(\varphi_1,\cdots,\varphi_N;\theta_1,\cdots,\theta_N;\tau).
\label{GREEN}
\end{equation} 
\par 
In this paper we focus on the case $\beta = 2$ , in which the interaction 
term in (\ref{CS}) vanishes, and calculate the dynamical correlation 
functions for typical initial conditions. Then $G^{(H)}$ is a determinant 
of one particle Green functions and ($\ref{GREEN}$) reads 
\begin{equation}
G(\varphi_1,\cdots,\varphi_N;\theta_1,\cdots,\theta_N;\tau) \propto 
\prod_{j>l}^N \frac{\displaystyle 
\sin\frac{\theta_j - \theta_l}{2}}{\displaystyle \sin\frac{\varphi_j 
- \varphi_l}{2}} {\rm det}[g(\theta_j,\varphi_l;\tau)]_{j,l=1,2,\cdots,N},
\end{equation}
where ($z = {\rm e}^{i \theta}, \ w = {\rm e}^{i \varphi}$)
\begin{equation}
g(\theta,\varphi;\tau) = \left\{ \begin{array}{ll} 
\displaystyle \frac{1}{2 \pi} \left( \frac{z}{w} \right)^{1/2}
\sum_{n=-\infty}^{\infty} \left( \frac{w}{z} \right)^n 
{\rm e}^{- \gamma_n \tau}, & N \ {\rm even}, \\
\displaystyle \frac{1}{2 \pi} \sum_{n=-\infty}^{\infty} 
\left( \frac{w}{z} \right)^n 
{\rm e}^{- \gamma_n \tau}, & N \ {\rm odd} 
\end{array} \right.
\end{equation}
and
\begin{equation}
\gamma_n = \left\{ \begin{array}{ll} (n-(1/2))^2/2, & N \ {\rm even}, \\ 
n^2/2, & N \ {\rm odd}. \end{array} \right.
\label{GAMMA}
\end{equation}
We remark that the following discussion does not depend on 
the particular form (\ref{GAMMA}) of $\gamma_n$, provided they 
are positive. 
\par
Let us suppose that the initial distribution of the 
eigenvalues is one of the followings (note that $U(\varphi)$ 
can be a function with a complex value):
\begin{equation}
p_0(\varphi_1,\cdots,\varphi_N) \propto 
\left\{ \begin{array}{lll}
\prod_{j=1}^N U(\varphi_j) 
\prod_{j<l}^N \mid {\rm e}^{i \varphi_j} - {\rm e}^{i \varphi_l} \mid, 
\\ \\ \sum_P \prod_{j=1}^{N/2} U(\varphi_j)^2 \delta(\varphi_j - 
\varphi_{j+N/2}) 
\\ \times \prod_{j<l}^{N/2} \mid {\rm e}^{i \varphi_j} 
- {\rm e}^{i \varphi_l} 
\mid^4,  \ \ \ ({\rm assuming} \ N \ {\rm even}), \\ \\ 
\sum_P \prod_{j=1}^{[N/2]} U(\varphi_j)^2 \delta(\varphi_j 
+ \varphi_{j + [(N+1)/2]}) \\ 
\times \prod_{j<l}^{[N/2]} 
\mid {\rm e}^{i \varphi_j} - {\rm e}^{i \varphi_l} \mid^2    
\mid {\rm e}^{i \varphi_j} - {\rm e}^{-i \varphi_l} \mid^2 
\\ \times \displaystyle \left\{ \begin{array}{ll} 1, 
\ \ \ N \ {\rm even}, \\ \delta(\varphi_{[(N+1)/2]}) \prod_{j=1}^{[N/2]} 
\mid 1  - {\rm e}^{i \varphi_j} \mid^2, \ \ \ N \ {\rm odd}.
\end{array} \right. \end{array} \right.
\label{INITIAL}
\end{equation}
Here we have introduced a sum $\sum_P$ over the permutation 
of $\varphi_j$'s in order to make the initial distribution totally 
symmetric. $[x]$ is the largest integer not exceeding $x$. 
The first and second of these initial conditions can 
be derived as eigenvalue distributions of symmetric unitary 
and self-dual quaternion unitary random matrices, respectively. 
The third one, for which the weight function $U(\varphi)$ is assumed 
to be symmetrical about the origin, is an analogue of the eigenvalue 
distribution of antisymmetric hermitian random matrices.  
\par
The probability distribution function $p$ of the eigenvalues is 
calculated from the initial condition and the Green functions as
\begin{eqnarray}
& & p(\theta_1^1,\cdots,\theta_N^2;\tau_1;\theta_1^2,\cdots,\theta_N^2;\tau_2; 
\cdots;\theta_1^M,\cdots,\theta_N^M;\tau_M) \nonumber \\  
& = & \frac{1}{N!} \int_{-\pi}^{\pi} {\rm d}\theta_1^0 
\cdots \int_{-\pi}^{\pi} {\rm d}\theta_N^0 
p_0(\theta_1^0,\cdots,\theta_N^0) \prod_{l=1}^M 
G(\theta_1^{l-1},\cdots,\theta_N^{l-1}; 
\theta_1^l,\cdots,\theta_N^l;\tau_l - \tau_{l-1}), 
\nonumber \\ & & \tau_0 = 0.
\end{eqnarray} 
For $N$ even, performing the integration gives
\begin{eqnarray}
& & p(\theta_1^1,\cdots,\theta_N^2;\tau_1;\theta_1^2,\cdots,\theta_N^2;\tau_2; 
\cdots;\theta_1^M,\cdots,\theta_N^M;\tau_M) \nonumber \\  
& = & i^{-N(N-1)/2} \prod_{j=1}^N 
{\rm e}^{- i (N-1)\theta_j^M/2}  
\prod_{j>l}^N ({\rm e}^{i \theta_j^M} - {\rm e}^{i \theta_l^M}) 
\nonumber \\ & \times &  
{\rm Pf}[F_{jl}^{11}]_{j,l=1,2,\cdots,N} \prod_{k=1}^{M-1} \det[g_{jl}^{k+1 \ 
k}]_{j,l=1,2,\cdots,N}. \nonumber \\
\label{PEVEN} 
\end{eqnarray}   
Here ${\rm Pf}$ means a Pfaffian. The matrices $g^{mn}$ 
and $F^{mn}$ are defined as
\begin{equation}
g_{jl}^{mn} = g(\theta_j^m,\theta_l^n;\tau_m-\tau_n)
\end{equation}
and
\begin{equation}
F_{jl}^{mn} = F(\theta_j^m,\theta_l^n;\tau_m,\tau_n),
\end{equation}
where
\begin{eqnarray}
& & F(\theta,\theta^{\prime};\tau,\tau^{\prime}) 
\nonumber \\ & = & 
\int_{-\pi}^{\pi} {\rm d}\varphi^{\prime} 
\int_{-\pi}^{\varphi^{\prime}} {\rm d}\varphi 
U(\varphi) U(\varphi^{\prime}) 
\{ g(\theta,\varphi;\tau) g(\theta^{\prime},
\varphi^{\prime};\tau^{\prime}) 
- g(\theta^{\prime},\varphi;\tau^{\prime}) 
g(\theta,\varphi^{\prime};\tau) \}, \nonumber \\ 
& & F(\theta,\theta^{\prime};\tau,\tau^{\prime}) 
\nonumber \\ & = & 
\int_{-\pi}^{\pi} {\rm d}\varphi 
U(\varphi)^2 
\{ g(\theta,\varphi;\tau) 
\frac{\partial}{\partial \varphi}g(\theta^{\prime},
\varphi;\tau^{\prime}) 
- g(\theta^{\prime},\varphi;\tau^{\prime}) 
\frac{\partial}{\partial \varphi}g(\theta,\varphi;\tau) \}, \nonumber \\ 
& & F(\theta,\theta^{\prime};\tau,\tau^{\prime}) 
\nonumber \\ & = & 
\int_0^{\pi} {\rm d}\varphi 
U(\varphi)^2 \frac{1}{2 \sin\varphi} 
\{ g(\theta,-\varphi;\tau) g(\theta^{\prime}, \varphi;\tau^{\prime}) 
- g(\theta^{\prime},-\varphi;\tau^{\prime}) 
g(\theta,\varphi;\tau) \} \nonumber \\ 
\label{FTHETA}
\end{eqnarray}
for each of the three initial conditions (\ref{INITIAL}), 
respectively.  For $N$ odd, performing the integration similarly 
yields 
\begin{eqnarray}
& & p(\theta_1^1,\cdots,\theta_N^2;\tau_1;\theta_1^2,\cdots,\theta_N^2;\tau_2; 
\cdots;\theta_1^M,\cdots,\theta_N^M;\tau_M) \nonumber \\  
& = & i^{-N(N-1)/2} \prod_{j=1}^N 
{\rm e}^{- i (N-1)\theta_j^M/2}  
\prod_{j>l}^N ({\rm e}^{i \theta_j^M} - {\rm e}^{i \theta_l^M}) \nonumber \\ 
& \times & {\rm Pf}\left[ \begin{array}{cc} [F_{jl}^{11}]_{j,l=1,2,\cdots,N} & 
[f^1_j]_{j=1,\cdots,N} \\ -[f^1_l]_{l=1,\cdots,N} & 0 \end{array} 
\right] \prod_{k=1}^{M-1} \det[g_{jl}^{k+1 \ 
k}]_{j,l=1,2,\cdots,N}. \nonumber \\ 
\label{PODD}
\end{eqnarray}
Here
\begin{equation}
f^m_j = f(\theta^m_j;\tau_m)
\end{equation}
with
\begin{equation}
f(\theta;\tau) = \int_{-\pi}^{\pi} U(\varphi) 
g(\theta,\varphi;\tau) {\rm d}\varphi, \ \ \ 
f(\theta;\tau) = U(0) g(\theta,0;\tau),
\end{equation}
for the first and last of the initial conditions (\ref{INITIAL}), 
respectively. 
\par 
Now we are in a position to define multilevel dynamical correlation 
functions 
\begin{eqnarray}
& & \rho(\theta^1_1,\cdots,\theta^1_{m_1}; \tau_1; 
\theta^2_1,\cdots,\theta^2_{m_2}; \tau_2; 
\cdots;\theta^M_1,\cdots,\theta^M_{m_M}; \tau_M) 
\nonumber \\ & = & 
\frac{1}{C_N} \frac{(N!)^M}{\prod_{l=1}^M (N-m_l)! } 
\int_{-\pi}^{\pi} {\rm d}\theta^1_{m_1+1} \cdots 
\int_{-\pi}^{\pi} {\rm d}\theta^1_N \cdots
\int_{-\pi}^{\pi} {\rm d}\theta^M_{m_M+1} 
\cdots \int_{-\pi}^{\pi} {\rm d}\theta^M_N \nonumber \\ 
& \times & p(\theta^1_1,\cdots,\theta^1_N;\tau_1; 
\theta^2_1,\cdots,\theta^2_N; \tau_2; 
\cdots; \theta^M_1,\cdots,\theta^M_N; \tau_M), \nonumber \\ 
\end{eqnarray}
where the normalization constant $C_N$ is defined as 
\begin{eqnarray}
C_N & = & 
\int_{-\pi}^{\pi} 
{\rm d}\theta^1_1 \cdots \int_{-\pi}^{\pi} {\rm d}\theta^1_N  \cdots 
\int_{-\pi}^{\pi} {\rm d}\theta^M_1 \cdots \int_{-\pi}^{\pi} {\rm d}\theta^M_N \nonumber \\ 
& \times & p(\theta^1_1,\cdots,\theta^1_N;\tau_1; 
\theta^2_1,\cdots,\theta^2_N; \tau_2; 
\cdots; 
\theta^M_1,\cdots,\theta^M_N; \tau_M).
\end{eqnarray}
Let us review the history of the study on multilevel 
correlation functions of random matrices and explain the 
purpose of this paper. The quaternion determinant 
formulas for the static correlations ($M=1,\tau_1=0$) with $U(\theta) = 1$ 
was discovered by Dyson\cite{DYQ}. His result was generalized to the case 
with a real weight function $U(\theta)$ by Nagao and Wadati\cite{NW}. 
The dynamical correlations were first explored for hermitian random 
matrix models by Pandey and Mehta \cite{PM,MP} and then extended to the unitary 
matrix case with $U(\theta)=1$ by Pandey 
and Shukla\cite{PS}. They were able to derive quaternion determinant forms 
for equal time ($M=1,0 \leq \tau_1 < \infty$) correlation functions. 
Recently Nagao and Forrester succeeded in evaluating the dynamical 
correlation functions with general $M$ for hermitian random 
matrices\cite{NFQ,TN}. 
In this paper we deal with unitary random matrices and generalize 
both of Nagao and Wadati's and Pandey and Shukla's results to show 
how the multilevel dynamical correlation functions with general $M$ 
can be written in quaternion determinant forms.  

\section{Dynamical Correlation Functions}
\setcounter{equation}{0}
\renewcommand{\theequation}{2.\arabic{equation}}

\subsection{Quaternion Determinant}

Let us begin with an introduction of a quaternion determinant, 
a determinant of a matrix with quaternion elements\cite{DYQ,MOORE1,MOORE2,
DYHELV,MEHTA2}. We define a quaternion as a linear combination of 
four basic units $\{1, e_1, e_2, e_3 \}$:
\begin{equation} 
q=q_0+{\bf q} \cdot {\bf e}=q_0+q_1e_1+q_2e_2+q_3e_3. 
\end{equation}
Here $q_0, q_1, q_2$ and $q_3$ are real or complex numbers. 
The first part $q_1$ is called the scalar part of $q$. 
The quaternion multiplication is associative 
but in general not commutative: the multiplication rule of 
the four basic units are
\begin{equation} 
1 \cdot 1=1,\;\; 1 \cdot e_j=e_j \cdot 
1=e_j,\;\; j=1,2,3, \nonumber 
\end{equation}
\begin{equation} 
e_1^2=e_2^2=e_3^2=e_1e_2e_3=-1. 
\end{equation} 
A quaternion $q$ has a dual ${\hat q}$ defined as 
\begin{equation} 
{\hat q} = q_0-{\bf q} \cdot {\bf e}. 
\end{equation} 
A matrix $Q$ with quaternion elements $q_{jl}$ also has a dual 
matrix ${\hat Q}=[{\hat q}_{lj}]$. The quaternion units can be  
represented as $2 \times 2$ matrices
\begin{displaymath} 
1 \rightarrow 
\left[ \begin{array}{cc} 1 & 0 \\ 0 & 1 \end{array} 
\right], \ \ e_1 \rightarrow \left[ \begin{array}{cc} 0 & -1 \\ 1 & 0 \end{array} \right],
\end{displaymath}
\begin{equation} 
e_2 \rightarrow \left[ \begin{array}{cc} 0 & -i \\ -i & 0 \end{array} \right], 
\ \ e_3 \rightarrow \left[ \begin{array}{cc} i & 0 \\ 0 & -i 
\end{array} \right]. 
\end{equation} 
\par
Now we are in a position to define a quaternion determinant {\rm Tdet}. 
For a self-dual $Q$ ( i.e., $Q={\hat Q}$ ), it is written as 
\begin{equation} {\rm Tdet}\, Q = \sum_P (-1)^{N-l} \prod_1^l (q_{ab} 
q_{bc} \cdots q_{da})_0.
\end{equation}
Here $P$ denotes any permutation of the indices 
$(1,2,\cdots,N)$ consisting of $l$ exclusive cycles of the form 
$(a \rightarrow b \rightarrow c \rightarrow \cdots \rightarrow d 
\rightarrow a)$. Note that $(-1)^{N-l}$ is the parity of $P$. 
The subscript $0$ has a meaning that we take the scalar part of 
the product over each cycle. If all the elements of $Q$ are scalars, 
then everything is commutable and a quaternion determinant becomes 
an ordinary determinant.

\subsection{The Case $N$ Even}
Throughout this paper we adopt a notation that 
${\overline z}$ is the complex conjugate of $z$ (note 
that $z^*$ is not necessarily the complex conjugate 
of $z$). Let us define ($z = {\rm e}^{i \theta}$, 
$w = {\rm e}^{i \varphi}$)
\begin{eqnarray}
\Psi^*_n(\theta;\tau) & = & 
\overline{\sqrt{z}} R^*_n(\theta;\tau), \nonumber \\ 
\Psi_n(\theta;\tau) & = & 
\sqrt{z} R_n(\theta;\tau), \nonumber \\ 
\Phi^*_n(\theta;\tau) & = & \int_{-\pi}^{\pi} 
F(\varphi,\theta;\tau,\tau) 
\overline{\sqrt{w}} 
R^*_n(\varphi;\tau) {\rm d}\varphi, \nonumber \\ 
\Phi_n(\theta;\tau) & = & \int_{-\pi}^{\pi} F(\varphi,\theta;\tau,\tau) 
\sqrt{w} R_n(\varphi;\tau) {\rm d}\varphi,
\end{eqnarray}
where $R^*_n(\theta;\tau) 
= z^{-n} + c^*_{n \ n-1} z^{-n+1} + \cdots 
+ c^*_{n \ -n} z^{n}$ and 
$R_n(\theta;\tau) = z^n + c_{n \ n-1} z^{n-1} + 
\cdots + c_{n \ -n} z^{-n}$ 
are arbitrary polynomials ($R^*_0(\theta;\tau) = 
R_0(\theta;\tau)=1$). 
\par
We introduce matrices 
$D^{mn}$, $I^{mn}$ and $S^{mn}$ as
\begin{equation}
D_{jl}^{mn} = D(\theta^m_j,\theta^n_l;\tau_m,\tau_n), \ \ \  
I_{jl}^{mn} = I(\theta^m_j,\theta^n_l;\tau_m,\tau_n), \ \ \  
S_{jl}^{mn} = S(\theta^m_j,\theta^n_l;\tau_m,\tau_n) 
\end{equation} 
with
\begin{eqnarray}
& & D(\theta,\theta^{\prime};\tau,\tau^{\prime}) \nonumber \\ 
& = &  
\sum_{k=0}^{(N/2)-1} \frac{1}{r_k(\tau^{\prime})} 
[ {\rm e}^{- \gamma_{-k}(\tau^{\prime} - \tau)} \Psi^*_k(\theta;\tau) 
\Psi_k(\theta^{\prime};\tau^{\prime}) - {\rm e}^{- \gamma_{k+1} 
(\tau^{\prime} - \tau)} \Psi_k(\theta;\tau) 
\Psi^*_k(\theta^{\prime};\tau^{\prime}) ], \nonumber \\ 
\end{eqnarray}
\begin{eqnarray}
& & I(\theta,\theta^{\prime};\tau,\tau^{\prime}) \nonumber \\ 
& = &  
- \sum_{k=0}^{(N/2)-1} \frac{1}{r_k(\tau^{\prime})} 
[ {\rm e}^{- \gamma_{-k}(\tau^{\prime} - \tau)} \Phi^*_k(\theta;\tau) 
\Phi_k(\theta^{\prime};\tau^{\prime}) - {\rm e}^{- \gamma_{k+1} 
(\tau^{\prime} - \tau)} 
\Phi_k(\theta;\tau) \Phi^*_k(\theta^{\prime};\tau^{\prime}) ] 
\nonumber \\ & + & F(\theta,\theta^{\prime};\tau,\tau^{\prime}) 
\end{eqnarray}
and
\begin{eqnarray}
& & S(\theta,\theta^{\prime};\tau,\tau^{\prime}) \nonumber \\ 
& = &  
\sum_{k=0}^{(N/2)-1} \frac{1}{r_k(\tau^{\prime})} 
[ {\rm e}^{- \gamma_{-k}(\tau^{\prime} - \tau)} \Phi^*_k(\theta;\tau) 
\Psi_k(\theta^{\prime};\tau^{\prime}) - {\rm e}^{- \gamma_{k+1} 
(\tau^{\prime} - \tau)} \Phi_k(\theta;\tau) 
\Psi^*_k(\theta^{\prime};\tau^{\prime}) ]. \nonumber \\ 
\label{STHETA}
\end{eqnarray}
Moreover we define
\begin{equation}
{\tilde S}_{jl}^{mn} = \left\{ \begin{array}{ll} 
S_{jl}^{mn} - g_{jl}^{mn}, \ \ \ m>n,\\ S^{mn}_{jl}, \ \ \ m \leq n. 
\end{array} \right.
\end{equation}
Then we have the following theorem.
\par
\bigskip
\noindent
{\em Theorem 1}
\par
\bigskip
\noindent
The probability distribution function (\ref{PEVEN}) with $N$ even 
can be written as a quaternion determinant
\begin{eqnarray}
& & p(\theta_1^1,\cdots,\theta_N^2;\tau_1;\theta_1^2,\cdots,\theta_N^2;\tau_2; 
\cdots;\theta_1^M,\cdots,\theta_N^M;\tau_M) 
\nonumber \\ & = & i^{-N(N-1)/2} 
\prod_{j=0}^{(N/2)-1} r_j(\tau_M) {\rm Tdet}[B^{\mu \nu}], \nonumber \\ 
& & \mu,\nu=1,2,\cdots,M.
\end{eqnarray}   
Each block $B^{\mu \nu}$ is an $N \times N$ quaternion matrix. Its 
quaternion elements are represented as 
\begin{equation}
B^{\mu \nu}_{jl} = 
\left[ \begin{array}{cc} {\tilde S}^{\mu \nu}_{jl} & 
I^{\mu \nu}_{jl} \\ 
D^{\mu \nu}_{jl} & {\tilde S}^{\nu \mu}_{lj} 
\end{array} \right]. 
\end{equation}
\par
\medskip
\noindent
Starting from (\ref{PEVEN}), we can follow a similar argument as in 
Refs.\cite{NFQ,TN} to prove {\em Theorem 1}.
\par
\medskip  
By Schmidt's orthogonalization procedure, 
we can specify the polynomials $R_n(\theta;\tau)$ 
and $R^*_n(\theta,\tau)$ so that they  satisfy 
the following skew orthogonality relation:
\begin{displaymath} \langle 
\overline{\sqrt{z}} R^*_m(\theta;\tau), \sqrt{w} 
R_n(\varphi;\tau) \rangle = 
- \langle \sqrt{z} R_n(\theta;\tau), 
\overline{\sqrt{w}} R^*_m(\varphi;\tau) \rangle  = 
r_m(\tau) \delta_{mn}, \end{displaymath}
\begin{displaymath} \langle \overline{\sqrt{z}} R^*_m(\theta;\tau), 
\overline{\sqrt{w}} 
R^*_n(\varphi;\tau) \rangle = 0, \quad
 \langle \sqrt{z}  R_m(\theta;\tau), \sqrt{w} 
R_n(\varphi;\tau) \rangle  
= 0,\end{displaymath}
\begin{equation}  \label{EVENSKEW}\end{equation}
where
\begin{equation}
\langle f(\theta), g(\varphi) \rangle = 
\int_{-\pi}^{\pi}  {\rm d}\theta 
\int_{-\pi}^{\pi} {\rm d}\varphi 
F(\theta,\varphi;\tau,\tau) 
f(\theta) g(\varphi).
\label{SKEW}
\end{equation}
Explicit determinant formula for the skew orthogonal 
polynomials $R_n(\theta;\tau)$ and $R^*(\theta,\tau)$ 
are given by ($n \geq 1$)\cite{NW}
\begin{equation}
R_n(z) = {\cal D}_n^{-1} \left| \begin{array}{ccccccc} 
z^n & J^{n \ n-1} & \cdots & J^{n \ 0} & J^{n \ -1} & \cdots & J^{n \ -n} \\ 
z^{n-1} & J^{n-1 \ n-1} & \cdots & J^{n-1 \ 0} & J^{n-1 \ -1} & \cdots & 
J^{n-1 \ -n} \\ 
\vdots & \vdots & \ddots & \vdots & \vdots & \ddots & \vdots \\ 
z^0 & J^{0 \ n-1} & \cdots & J^{0 \ 0} & J^{0 \ -1} & \cdots & J^{0 \ -n} \\ 
\vdots & \vdots & \ddots & \vdots & \vdots & \ddots & \vdots \\ 
z^{-n} & J^{-n \ n-1} & \cdots & J^{-n \ 0} & J^{-n \ -1} & \cdots & 
J^{-n \ -n} \end{array} \right|
\end{equation}
and 
\begin{equation}
R^*_n(z) = {\cal D}_n^{-1} \left| \begin{array}{ccccccc} 
z^{-n} & J^{-n-1 \ -n} & \cdots & J^{-n-1 \ 0} & J^{-n-1 \ 1} 
& \cdots & J^{-n-1 \ n-1} \\ 
z^{-n+1} & J^{-n \ -n} & \cdots & J^{-n \ 0} & J^{-n \ 1} & \cdots & 
J^{-n \ n-1} \\ 
\vdots & \vdots & \ddots & \vdots & \vdots & \ddots & \vdots \\ 
z^0 & J^{-1 \ -n} & \cdots & J^{-1 \ 0} & J^{-1 \ 1} & \cdots & J^{-1 \ n-1} \\ 
\vdots & \vdots & \ddots & \vdots & \vdots & \ddots & \vdots \\ 
z^n & J^{n-1 \ -n} & \cdots & J^{n-1 \ 0} & J^{n-1 \ 1} & \cdots & 
J^{n-1 \ n-1} \end{array} \right|,
\end{equation}
where
\begin{equation}
J^{mn} = \langle z^{m+(1/2)},w^{n+(1/2)} \rangle
\end{equation}
and
\begin{equation}
{\cal D}_n =  \left| \begin{array}{cccccc} 
J^{n-1 \ n-1} & \cdots & J^{n-1 \ 0} & J^{n-1 \ -1} & \cdots & 
J^{n-1 \ -n} \\ 
\vdots & \ddots & \vdots & \vdots & \ddots & \vdots \\ 
J^{0 \ n-1} & \cdots & J^{0 \ 0} & J^{0 \ -1} & \cdots & J^{0 \ -n} \\ 
\vdots & \ddots & \vdots & \vdots & \ddots & \vdots \\ 
J^{-n \ n-1} & \cdots & J^{-n \ 0} & J^{-n \ -1} & \cdots & 
J^{-n \ -n} \end{array} \right|.
\end{equation}
\par
\subsection{The Case $N$ Odd}
Let us next consider the case with $N$ odd. We define 
\begin{eqnarray}
\Phi^*_n(\theta;\tau) & = & \int_{-\pi}^{\pi} 
F(\varphi,\theta;\tau,\tau) 
R^*_n(\varphi;\tau) {\rm d}\varphi, \nonumber \\ 
\Phi_n(\theta;\tau) & = & \int_{-\pi}^{\pi} F(\varphi,\theta;\tau,\tau) 
R_n(\varphi;\tau) {\rm d}\varphi.
\end{eqnarray}
Here $R_n(\theta;\tau)$ and $R^*_n(\theta;\tau)$ 
are arbitrary functions provided that
\begin{eqnarray}
\prod_{j>l}^N ( {\rm e}^{i \theta_j} - {\rm e}^{i \theta_l} ) & = & 
\prod_{j=1}^N {\rm e}^{i(N-1)\theta_j/2} \nonumber \\ & \times &  
\left| \begin{array}{cccc} 
R^*_{(N-1)/2}(\theta_1;\tau) & R^*_{(N-1)/2}(\theta_2;\tau)   
& \cdots & R^*_{(N-1)/2}(\theta_N;\tau) \\   
\vdots & \vdots & \ddots & \vdots  \\  
R^*_1(\theta_1;\tau) & R^*_1(\theta_2;\tau)   
& \cdots & R^*_1(\theta_N;\tau) \\   
R_0(\theta_1;\tau) & R_0(\theta_2;\tau)   
& \cdots & R_0(\theta_N;\tau) \\   
R_1(\theta_1;\tau) & R_1(\theta_2;\tau)   
& \cdots & R_1(\theta_N;\tau) \\   
\vdots & \vdots & \ddots & \vdots \\  
R_{(N-1)/2}(\theta_1;\tau) & R_{(N-1)/2}(\theta_2;\tau)   
& \cdots & R_{(N-1)/2}(\theta_N;\tau) 
\end{array} \right|. \nonumber \\ 
\label{RODD}
\end{eqnarray}  
Then matrices $D^{mn}$, $I^{mn}$ and $S^{mn}$ are introduced as 
\begin{equation}
D_{jl}^{mn} = D(\theta^m_j,\theta^n_l;\tau_m,\tau_n), \ \ \  
I_{jl}^{mn} = I(\theta^m_j,\theta^n_l;\tau_m,\tau_n), \ \ \  
S_{jl}^{mn} = S(\theta^m_j,\theta^n_l;\tau_m,\tau_n), 
\end{equation} 
where
\begin{eqnarray}
& & D(\theta,\theta^{\prime};\tau,\tau^{\prime}) \nonumber \\ 
& = &  
\sum_{k=1}^{(N-1)/2} \frac{1}{r_k(\tau^{\prime})} 
[ {\rm e}^{- \gamma_{-k}(\tau^{\prime} - \tau)} R^*_k(\theta;\tau) 
R_k(\theta^{\prime};\tau^{\prime}) - {\rm e}^{- \gamma_k 
(\tau^{\prime} - \tau)} R_k(\theta;\tau) 
R^*_k(\theta^{\prime};\tau^{\prime}) ], \nonumber \\ 
\end{eqnarray}
\begin{eqnarray}
& & I(\theta,\theta^{\prime};\tau,\tau^{\prime}) \nonumber \\ 
& = &  
- \sum_{k=1}^{(N-1)/2} \frac{1}{r_k(\tau^{\prime})} 
[ {\rm e}^{- \gamma_{-k}(\tau^{\prime} - \tau)} \Phi^*_k(\theta;\tau) 
\Phi_k(\theta^{\prime};\tau^{\prime}) - {\rm e}^{- \gamma_k 
(\tau^{\prime} - \tau)} 
\Phi_k(\theta;\tau) \Phi^*_k(\theta^{\prime};\tau^{\prime}) ] 
\nonumber \\ & + & 
\frac{1}{s_0(\tau)} \Phi_0(\theta;\tau) 
f(\theta^{\prime};\tau^{\prime}) - \frac{1}{s_0(\tau^{\prime})} 
\Phi_0(\theta^{\prime};\tau^{\prime}) f(\theta;\tau) + 
F(\theta,\theta^{\prime};\tau,\tau^{\prime}) \nonumber \\  
\end{eqnarray}
and
\begin{eqnarray}
& & S(\theta,\theta^{\prime};\tau,\tau^{\prime}) \nonumber \\ 
& = &  
\sum_{k=1}^{(N-1)/2} \frac{1}{r_k(\tau^{\prime})} 
[ {\rm e}^{- \gamma_{-k}(\tau^{\prime} - \tau)} \Phi^*_k(\theta;\tau) 
R_k(\theta^{\prime};\tau^{\prime}) - {\rm e}^{- \gamma_k 
(\tau^{\prime} - \tau)} \Phi_k(\theta;\tau) 
R^*_k(\theta^{\prime};\tau^{\prime}) ] \nonumber \\ 
& + & \frac{1}{s_0(\tau^{\prime})} R_0(\theta^{\prime},\tau^{\prime}) 
f(\theta;\tau).
\label{SODD}
\end{eqnarray}
As before we set
\begin{equation}
{\tilde S}_{jl}^{mn} = \left\{ \begin{array}{ll} 
S_{jl}^{mn} - g_{jl}^{mn}, \ \ \ m>n,\\ S^{mn}_{jl}, \ \ \ m \leq n 
\end{array} \right.
\end{equation}
and find the following theorem.
\par
\bigskip
\noindent
{\em Theorem 2}
\par
\bigskip
\noindent
The probability distribution function (\ref{PODD}) with $N$ odd 
can be rewritten as 
\begin{eqnarray}
& & p(\theta_1^1,\cdots,\theta_N^2;\tau_1;\theta_1^2,\cdots,\theta_N^2;\tau_2; 
\cdots;\theta_1^M,\cdots,\theta_N^M;\tau_M) 
\nonumber \\ & = & i^{-N(N-1)/2} s_0(\tau_M) 
\prod_{j=1}^{(N-1)/2} r_j(\tau_M) {\rm Tdet}[B^{\mu \nu}], \nonumber \\ 
& & \mu,\nu=1,2,\cdots,M.
\end{eqnarray}   
Each block $B^{\mu \nu}$ is an $N \times N$ quaternion matrix 
the elements of which are represented as 
\begin{equation}
B^{\mu \nu}_{jl} = 
\left[ \begin{array}{cc} {\tilde S}^{\mu \nu}_{jl} & 
I^{\mu \nu}_{jl} \\ 
D^{\mu \nu}_{jl} & {\tilde S}^{\nu \mu}_{lj} 
\end{array} \right]. 
\end{equation}
\par
\medskip
\noindent
{\em Theorem 2} can be proven from (\ref{PODD}) by proceeding as in 
Refs.\cite{NFQ,TN}.
\par
\medskip  
We now introduce a set of polynomials $\Omega_n(\theta;\tau)$, 
$\Omega^*_n(\theta;\tau)$ for $n \geq 1$ and $\Omega_0(\theta;\tau) 
= \Omega^*_0(\theta;\tau)$ 
satisfying 
\begin{displaymath} \langle 
\Omega^*_m(\theta;\tau),  
\Omega_n(\varphi;\tau) \rangle = 
- \langle \Omega_n(\theta;\tau), 
\Omega^*_m(\varphi;\tau) \rangle  = 
r_m(\tau) \delta_{mn}, \end{displaymath}
\begin{equation} \langle \Omega^*_m(\theta;\tau), 
\Omega^*_n(\varphi;\tau) \rangle = 0, \quad \langle 
\Omega_m(\theta;\tau),  
\Omega_n(\varphi;\tau) \rangle  
= 0 \label{ODDSKEW}\end{equation}
for $m,n \geq 1$ and
\begin{displaymath} \langle 
\Omega^*_n(\theta;\tau),  
\Omega_0(\varphi;\tau) \rangle = 
- \langle \Omega_0(\theta;\tau), 
\Omega^*_n(\varphi;\tau) \rangle  = 
0, \end{displaymath}
\begin{equation} \langle 
\Omega^*_0(\theta;\tau),  
\Omega_n(\varphi;\tau) \rangle = 
- \langle \Omega_n(\theta;\tau), 
\Omega^*_0(\varphi;\tau) \rangle  = 
0 \label{ZEROSKEW}\end{equation}
for $0 \leq n \leq (N-1)/2$.  Here the bracket is defined 
in (\ref{SKEW}). 
\par
Let us define
\begin{equation}
K^{mn} = \left\{ \begin{array}{ll} 
\langle z^m, \Omega_n(\varphi;\tau) \rangle, & n > 0, \\ 
\langle z^m, \Omega^*_n(\varphi;\tau) \rangle, & n < 0 
\end{array} \right.
\end{equation}
and
\begin{equation}
L^{mn} = \left\{ \begin{array}{ll} 
\langle \Omega_m(\theta;\tau), \Omega_n(\varphi;\tau) \rangle, & m,n > 0, \\ 
\langle \Omega_m(\theta;\tau), \Omega^*_n(\varphi;\tau) \rangle, & m > 0,n < 0, \\ 
\langle \Omega^*_m(\theta;\tau), \Omega_n(\varphi;\tau) \rangle, & m < 0, n > 0, \\ 
\langle \Omega^*_m(\theta;\tau), \Omega^*_n(\varphi;\tau) \rangle, & m,n < 0. 
\end{array} \right.
\end{equation}
Then, starting from ($z = {\rm e}^{i \theta}$) 
\begin{equation}
\Omega_1(\theta;\tau) = a_1 + z, \ \ \ \Omega_1^*(\theta;\tau) = a^*_1 + \frac{1}{z},
\end{equation}
we can recursively construct the polynomials as ($n \geq 2$)
\begin{eqnarray}
& & \Omega_n(\theta;\tau) = a_n \Omega^{(n-1)}_0(\theta;\tau)
\nonumber \\ 
& + & {\cal D}_{n-1}^{-1} \left| \begin{array}{ccccccc} 
z^n & K^{n \ n-1} & \cdots & K^{n \ 1} & K^{n \ -1} 
& \cdots & K^{n \ -n+1} \\ 
\Omega_{n-1}(\theta;\tau) & L^{n-1 \ n-1} 
& \cdots & L^{n-1 \ 1} & L^{n-1 \ -1} & \cdots & 
L^{n-1 \ -n+1} \\ 
\vdots & \vdots & \ddots & \vdots & \vdots & \ddots & \vdots \\ 
\Omega_1(\theta;\tau) & L^{1 \ n-1} & \cdots & L^{1 \ 1} & L^{1 \ - 1} 
& \cdots & L^{1 \ -n+1} \\ 
\Omega^*_1(\theta;\tau) & L^{-1 \ n-1} & \cdots & L^{-1 \ 1} & L^{-1 \ - 1} 
& \cdots & L^{-1 \ -n+1} \\ 
\vdots & \vdots & \ddots & \vdots & \vdots & \ddots & \vdots \\ 
\Omega^*_{n-1}(\theta;\tau) & L^{-n+1 \ n-1} & \cdots & L^{-n+1 \ 1} & L^{-n+1 \ -1} 
& \cdots & L^{-n+1 \ -n+1} \end{array} \right| \nonumber \\ 
\end{eqnarray}
and
\begin{eqnarray}
& & \Omega^*_n(\theta;\tau) = a^*_n \Omega^{(n-1)}_0(\theta;\tau)
\nonumber \\ 
& + & {\cal D}_{n-1}^{-1} \left| \begin{array}{ccccccc} 
z^{-n} & K^{-n \ -n+1} & \cdots & K^{-n \ -1} & K^{-n \ 1} 
& \cdots & K^{-n \ n-1} \\ 
\Omega^*_{n-1}(\theta;\tau) & L^{-n+1 \ -n+1} & \cdots & L^{-n+1 \ -1} 
& L^{-n+1 \ 1} & \cdots & 
L^{-n+1 \ n-1} \\ 
\vdots & \vdots & \ddots & \vdots & \vdots & \ddots & \vdots \\ 
\Omega^*_1(\theta;\tau) & L^{-1 \ -n+1} & \cdots & L^{-1 \ -1} & L^{-1 \ 1} 
& \cdots & L^{-1 \ n-1} \\ 
\Omega_1(\theta;\tau) & L^{1 \ -n+1} & \cdots & L^{1 \ -1} & L^{1 \ 1} 
& \cdots & L^{1 \ n-1} \\ 
\vdots & \vdots & \ddots & \vdots & \vdots & \ddots & \vdots \\ 
\Omega_{n-1}(\theta;\tau) & L^{n-1 \ -n+1} & \cdots & L^{n-1 \ -1} & L^{n-1 \ 1} 
& \cdots & L^{n-1 \ n-1} \end{array} \right|. \nonumber \\ 
\end{eqnarray}
Here
\begin{eqnarray}
& & \Omega^{(n-1)}_0(\theta;\tau)
\nonumber \\ 
& = & {\cal D}_{n-1}^{-1} \left| \begin{array}{ccccccc} 
1 & K^{0 \ n-1} & \cdots & K^{0 \ 1} & K^{0 \ -1} 
& \cdots & K^{0 \ -n+1} \\ 
\Omega_{n-1}(\theta;\tau) & L^{n-1 \ n-1} 
& \cdots & L^{n-1 \ 1} & L^{n-1 \ -1} & \cdots & 
L^{n-1 \ -n+1} \\ 
\vdots & \vdots & \ddots & \vdots & \vdots & \ddots & \vdots \\ 
\Omega_1(\theta;\tau) & L^{1 \ n-1} & \cdots & L^{1 \ 1} & L^{1 \ - 1} 
& \cdots & L^{1 \ -n+1} \\ 
\Omega^*_1(\theta;\tau) & L^{-1 \ n-1} & \cdots & L^{-1 \ 1} & L^{-1 \ - 1} 
& \cdots & L^{-1 \ -n+1} \\ 
\vdots & \vdots & \ddots & \vdots & \vdots & \ddots & \vdots \\ 
\Omega^*_{n-1}(\theta;\tau) & L^{-n+1 \ n-1} & \cdots & L^{-n+1 \ 1} & L^{-n+1 \ -1} 
& \cdots & L^{-n+1 \ -n+1} \end{array} \right| \nonumber \\ 
\end{eqnarray}
and
\begin{equation}
{\cal D}_{n-1} =  \left| \begin{array}{cccccc} 
L^{n-1 \ n-1} & \cdots & L^{n-1 \ 1} & L^{n-1 \ -1} & \cdots & 
L^{n-1 \ -n+1} \\ 
\vdots & \ddots & \vdots & \vdots & \ddots & \vdots \\ 
L^{1 \ n-1} & \cdots & L^{1 \ 1} & L^{1 \ - 1} 
& \cdots & L^{1 \ -n+1} \\ 
L^{-1 \ n-1} & \cdots & L^{-1 \ 1} & L^{-1 \ - 1} 
& \cdots & L^{-1 \ -n+1} \\ 
 \vdots & \ddots & \vdots & \vdots & \ddots & \vdots \\ 
L^{-n+1 \ n-1} & \cdots & L^{-n+1 \ 1} & L^{-n+1 \ -1} 
& \cdots & L^{-n+1 \ -n+1} \end{array} \right|. \nonumber \\ 
\end{equation}
Setting 
\begin{equation}
\Omega_0(\theta;\tau) = \Omega^{((N-1)/2)}_0(\theta;\tau),
\end{equation}
we find that all of the skew orthogonality 
conditions (\ref{ODDSKEW}) and (\ref{ZEROSKEW}) are satisfied. 
Note that there is an ambiguity in the determination of 
the skew orthogonal polynomials due to the constants 
$a_n$ and $a^*_n$.
\par
Let us now introduce ($n \geq 0$) 
\begin{eqnarray}
s^*_n(\tau) & = & \int_{-\pi}^{\pi} {\rm d}\theta f(\theta;\tau) 
\Omega^*_n(\theta;\tau), \nonumber \\ 
s_n(\tau) & = & \int_{-\pi}^{\pi} {\rm d}\theta f(\theta;\tau) 
\Omega_n(\theta;\tau).    
\end{eqnarray}
Then $R_n(\theta;\tau)$ and $R^*(\theta;\tau)$ 
satisfying (\ref{RODD}) 
can be constructed as ($n=1,2,\cdots,(N-1)/2$)
\begin{eqnarray}
R^*_n(\theta;\tau) & = & \Omega^*_n(\theta;\tau) - 
\frac{s^*_n(\tau)}{s_0(\tau)} \Omega_0(\theta;\tau), 
\nonumber \\  
R_n(\theta;\tau) & = & \Omega_n(\theta;\tau) - 
\frac{s_n(\tau)}{s_0(\tau)} \Omega_0(\theta;\tau) 
\end{eqnarray}
and 
\begin{equation}
R^*_0(\theta;\tau) = R_0(\theta;\tau) = \Omega_0(\theta;\tau).
\end{equation} 
\par
\subsection{Quaternion Determinant Expressions}
Using the orthogonality relations introduced above, it can 
be readily proven in both the cases $N$ even and odd that
\begin{eqnarray}
& & \int_{-\pi}^{\pi} B^{mp}_{ji} B^{pn}_{il} {\rm d}\theta^p_i \nonumber \\ 
& = & \left\{ \begin{array}{ll} 
0, & \ \ m < p, \ p > n, \\ 
B^{mn}_{jl} \left[ \begin{array}{cc} 1 & 0 \\ 0 & 0 \end{array} \right], & 
 \ \ m < p, \ p = n, \\ 
\left[ \begin{array}{cc} 0 & 0 \\ 0 & 1 \end{array} \right] B^{mn}_{jl}, & 
 \ \ m = p, \ p > n, \\ 
\left[ \begin{array}{cc} 0 & 0 \\ 0 & 1 \end{array} \right] B^{mn}_{jl} + 
B^{mn}_{jl} \left[ \begin{array}{cc} 1 & 0 \\ 0 & 0 \end{array} \right], & 
 \ \ m = p, \ p = n, \\ 
\left[ \begin{array}{cc} 0 & 0 \\ 0 & 1 \end{array} \right] B^{mn}_{jl} + 
B^{mn}_{jl} \left[ \begin{array}{cc} 1 & 0 \\ 0 & 0 \end{array} \right] - 
\left[ \begin{array}{cc} 1 & 0 \\ 0 & 0 \end{array} \right] B^{mn}_{jl} - 
B^{mn}_{jl} \left[ \begin{array}{cc} 0 & 0 \\ 0 & 1 \end{array} \right], & 
 \ \ m > p, \ p < n, \\ 
\left[ \begin{array}{cc} 0 & 0 \\ 0 & 1 \end{array} \right] B^{mn}_{jl} + 
B^{mn}_{jl} \left[ \begin{array}{cc} 1 & 0 \\ 0 & 0 \end{array} \right] - 
\left[ \begin{array}{cc} 1 & 0 \\ 0 & 0 \end{array} \right] B^{mn}_{jl}, & 
 \ \ m > p, \ p = n, \\ 
\left[ \begin{array}{cc} 0 & 0 \\ 0 & 1 \end{array} \right] B^{mn}_{jl} + 
B^{mn}_{jl} \left[ \begin{array}{cc} 1 & 0 \\ 0 & 0 \end{array} \right] -  
B^{mn}_{jl} \left[ \begin{array}{cc} 0 & 0 \\ 0 & 1 \end{array} \right], & 
 \ \ m = p, \ p < n, \\ 
\left[ \begin{array}{cc} 0 & 0 \\ 0 & 1 \end{array} \right] B^{mn}_{jl} - 
\left[ \begin{array}{cc} 1 & 0 \\ 0 & 0 \end{array} \right] B^{mn}_{jl}, & 
 \ \ m > p, \ p > n, \\ 
B^{mn}_{jl} \left[ \begin{array}{cc} 1 & 0 \\ 0 & 0 \end{array} \right] - 
B^{mn}_{jl} \left[ \begin{array}{cc} 0 & 0 \\ 0 & 1 \end{array} \right], & 
 \ \ m < p, \ p < n. 
\end{array} \right. \nonumber \\ 
\end{eqnarray}
Moreover we can easily find 
\begin{equation}
\int_{-\pi}^{\pi} B^{mm}_{jj} {\rm d}\theta^m_j = N.
\end{equation}
We then arrive at the following theorem which has the central 
importance in the evaluation of the dynamical correlations.
\par
\bigskip
\noindent
{\em Theorem 3}
\par
\bigskip
\noindent
In terms of the quaternion elements $B^{mn}_{jl}$, we define 
quaternion rectangular matrices
\begin{equation}
Q^{mn}_{JL} = \left( \begin{array}{ccc} B^{mn}_{11} & \cdots & B^{mn}_{1L} \\ 
\vdots & \ddots & \vdots \\ B^{mn}_{J1} & \cdots & B^{mn}_{JL} 
\end{array} \right).
\end{equation}
Let us further define a self-dual quaternion matrix $Q$ consisting of $Q^{mn}_{JL}$ as 
\begin{equation}
Q = \left( \begin{array}{ccc} Q^{11}_{n_1 \ n_1} & \cdots & Q^{1 K}_{n_1 \ n_K} \\ 
\vdots & \ddots & \vdots \\ Q^{K 1}_{n_K \ n_1} & \cdots & Q^{KK}_{n_K \ n_K} 
\end{array} \right).
\end{equation}
We denote the $j$-th row (column) of the $m$-th row (column) block of the 
quaternion matrix $Q$ as $(m,j)$ row (column). The $(m,j)$ row and $(m,j)$ column contain the 
variable $\theta^m_j$. If the $(m,j)$ row and $(m,j)$ column are removed, the resulting 
smaller matrix $Q^m_j$ is still self-dual and does not contain 
the variable $\theta^m_j$. 
\par
Using the above definitions, we have 
\begin{equation}
\int_{-\pi}^{\pi} {\rm d} \theta^{\mu}_{n_{\mu}} {\rm Tdet} Q = (N - n_{\mu} + 1) {\rm Tdet} Q^{\mu}_{n_{\mu}}.
\end{equation}
\par
\medskip
\noindent
The proof of {\em Theorem 3} is found by following the strategy 
given in Ref.\cite{NFQ}.
\par
\medskip
Successive application of {\em Theorem 3} 
leads to the quaternion determinant expression 
for the multilevel dynamical correlation function as 
\begin{eqnarray}
& & \rho(\theta^1_1,\cdots,\theta^1_{m_1}; 
\tau_1; \theta^2_1,\cdots,\theta^2_{m_2}; 
\tau_2; \cdots;\theta^M_1,\cdots,\theta^M_{m_M}; 
\tau_M)  = {\rm Tdet}[B^{\mu \nu}(m_{\mu}, m_{\nu})], \nonumber \\ 
& & \mu,\nu = 1,2,\cdots,M,
\end{eqnarray}
where each block $B^{\mu \nu}(m_{\mu}, m_{\nu}) $ is obtained by removing the $m_{\mu} + 1, m_{\mu} + 2,  \cdots, N$-th rows and $m_{\nu} + 1, m_{\nu} + 2, \cdots, N$-th columns from $B^{\mu \nu}$.  

\section{Dynamical Correlation within the Unitary Symmetry}
\setcounter{equation}{0}
\renewcommand{\theequation}{3.\arabic{equation}}
In this section we consider the limit $\tau_j \rightarrow \infty$, $j=1,2,\cdots,M$ with all $\tau_j - \tau_l$, $j,l = 1,2,\cdots,M$ fixed. In this limit it is 
expected that the dynamical correlations describe the Brownian dynamics within the unitary symmetry. We can conveniently take this limit by using the following summation formulas.
  
\subsection{The Case $N$ Even}
Let us express the skew orthogonal polynomials with $\tau = 0$ as 
($z = {\rm e}^{i \theta}$)
\begin{eqnarray}
R_n(\theta;0) & = & \sum_{j=-n}^n \alpha_{nj} z^j, \ \ \ \alpha_{nn} = 1, 
\nonumber \\ 
R^*_n(\theta;0) & = & \sum_{j=-n}^n \alpha^*_{nj} z^{-j}, \ \ \ 
\alpha^*_{nn} = 1.
\end{eqnarray}
Then it can be seen that
\begin{eqnarray}
R_n(\theta;\tau) & = & 
{\rm e}^{-\gamma_{n+1} \tau} 
\sum_{j=-n}^n \alpha_{nj} z^j {\rm e}^{\gamma_{j+1} \tau}, 
\nonumber \\ 
R^*_n(\theta;\tau) & = & 
{\rm e}^{-\gamma_{-n} \tau} 
\sum_{j=-n}^n \alpha^*_{nj} z^{-j} {\rm e}^{\gamma_{-j} \tau},
\label{REVENS} 
\end{eqnarray}
with
\begin{equation}
r_n(\tau)  = {\rm e}^{-\gamma_{-n} \tau} {\rm e}^{-\gamma_{n+1} \tau} 
r_n(0),
\end{equation}
the proof of which comes from an identity
\begin{equation}
\langle \overline{\sqrt{z}} 
R^*_m(\theta;\tau), \sqrt{w} R_n(\varphi;\tau) 
\rangle
= {\rm e}^{-\gamma_{-m} \tau} {\rm e}^{-\gamma_{n+1} \tau} 
\left. \langle \overline{\sqrt{z}} 
R^*_m(\theta;0), \sqrt{w} R_n(\varphi;0) 
\rangle \right|_{\tau=0}.
\end{equation}
Let us introduce an inverse expansion of (\ref{REVENS}) ($n \geq 1$) 
\begin{eqnarray}
z^n {\rm e}^{\gamma_{n+1} \tau} & = &  \sum_{j=0}^n \lambda_{nj} 
{\rm e}^{\gamma_{j+1} \tau} R_j(\theta;\tau) + \frac{1}{z} \sum_{j=0}^{n-1} 
\mu_{nj} {\rm e}^{\gamma_{-j} \tau} R^*_j(\theta;\tau),
\nonumber \\ 
z^{-n} {\rm e}^{\gamma_{-n} \tau} & = &  \sum_{j=0}^n \lambda^*_{nj} 
{\rm e}^{\gamma_{-j} \tau} R^*_j(\theta;\tau) + z \sum_{j=0}^{n-1} 
\mu^*_{nj} {\rm e}^{\gamma_{j+1} \tau} R_j(\theta;\tau). 
\label{ZEVENS}
\end{eqnarray}
Using (\ref{SEVENS}) and (\ref{ZEVENS}), we can readily derive 
($\lambda_{jj}=1, \ n \geq 0$) 
\begin{eqnarray}
\Phi^*_n(\theta;\tau) & = & \frac{{\rm e}^{- \gamma_{-n} \tau}}{2 \pi} 
r_n(0) \left[ 
\sum_{\nu \geq n}^{\infty} 
z^{-\nu - (1/2)} {\rm e}^{-\gamma_{\nu+1} 
\tau} \lambda_{\nu n} + 
\sum_{\nu \geq n + 1}^{\infty} 
z^{\nu + (1/2)} {\rm e}^{-\gamma_{-\nu} \tau} 
\mu^*_{\nu n} \right], \nonumber \\ 
\Phi_n(\theta;\tau) & = & - \frac{{\rm e}^{- \gamma_{n+1} \tau}}{2 \pi} 
r_n(0) \left[ 
\sum_{\nu \geq n + 1}^{\infty} 
z^{-\nu - (1/2)} {\rm e}^{-\gamma_{\nu+1} 
\tau} \mu_{\nu n} + 
\sum_{\nu \geq n}^{\infty} 
z^{\nu + (1/2)} {\rm e}^{-\gamma_{-\nu} \tau} 
\lambda^*_{\nu n} \right]. \nonumber \\ 
\label{PEVENS}
\end{eqnarray} 
Putting (\ref{ZEVENS}) into (\ref{FTHETA}) and comparing the 
result with (\ref{PEVENS}) lead to
\begin{eqnarray}
& & F (\theta, \theta^{\prime}; \tau, \tau^{ \prime}) \nonumber \\ 
& = &  
\sum_{k=0}^{\infty} \frac{1}{r_k(\tau^{\prime})} 
[ {\rm e}^{- \gamma_{-k}(\tau^{\prime} - \tau)} \Phi^*_k(\theta;\tau) 
\Phi_k(\theta^{\prime};\tau^{\prime}) - {\rm e}^{- \gamma_{k+1} 
(\tau^{\prime} - \tau)} 
\Phi_k(\theta;\tau) \Phi^*_k(\theta^{\prime};\tau^{\prime}) ], 
\nonumber \\ 
\end{eqnarray}
so that
\begin{eqnarray}
& & I(\theta,\theta^{\prime};\tau,\tau^{\prime}) \nonumber \\ 
& = &  
\sum_{k=N/2}^{\infty} \frac{1}{r_k(\tau^{\prime})} 
[ {\rm e}^{- \gamma_{-k}(\tau^{\prime} - \tau)} \Phi^*_k(\theta;\tau) 
\Phi_k(\theta^{\prime};\tau^{\prime}) - {\rm e}^{- \gamma_{k+1} 
(\tau^{\prime} - \tau)} 
\Phi_k(\theta;\tau) \Phi^*_k(\theta^{\prime};\tau^{\prime}) ].  
\nonumber \\ 
\end{eqnarray}
Substitution of (\ref{PEVENS}) into (\ref{STHETA}) results in
\begin{equation}
S(\theta,\theta^{\prime};\tau,\tau^{\prime}) =  
S_1(\theta,\theta^{\prime};\tau,\tau^{\prime}) 
+ S_2(\theta,\theta^{\prime};\tau,\tau^{\prime}),
\label{SEVENS}
\end{equation}
where ($z = {\rm e}^{i \theta}, \ z^{\prime} = 
{\rm e}^{i \theta^{\prime}}$)
\begin{equation}
S_1(\theta,\theta^{\prime};\tau,\tau^{\prime}) = 
\frac{1}{2 \pi} \left( \frac{z}{z^{\prime}} \right)^{1/2} 
\sum_{n=-(N/2)+1}^{N/2} \left( \frac{z^{\prime}}{z} \right)^n 
{\rm e}^{-\gamma_n (\tau - \tau^{\prime})}
\end{equation}
and
\begin{eqnarray}
& & S_2(\theta,\theta^{\prime};\tau,\tau^{\prime}) 
= \frac{1}{2 \pi}  
\sum_{k=0}^{(N/2)-1} 
\sum_{\nu=N/2}^{\infty} \nonumber \\ 
& \times & \left[ \left(z^{-\nu-(1/2)} {\rm e}^{-\gamma_{\nu+1} \tau} 
\lambda_{\nu k} + z^{\nu+(1/2)} {\rm e}^{-\gamma_{-\nu} \tau} 
\mu^*_{\nu k}) {\rm e}^{\gamma_{k+1} \tau^{\prime}} \sqrt{z^{\prime}} 
R_k(\theta^{\prime};\tau^{\prime}\right) \right. \nonumber \\ 
& +  & \left. \left(z^{-\nu-(1/2)} {\rm e}^{-\gamma_{\nu+1} \tau} 
\mu_{\nu k} + z^{\nu+(1/2)} {\rm e}^{-\gamma_{-\nu} \tau} 
\lambda^*_{\nu k}\right) {\rm e}^{\gamma_{-k} \tau^{\prime}} 
\overline{\sqrt{z^{\prime}}} 
R^*_k(\theta^{\prime};\tau^{\prime}) \right]. \nonumber \\ 
\end{eqnarray}
\par
Let us assume that $\gamma_n=\gamma_{-n+1}$ and $\gamma_{n+1} > \gamma_n$ 
for $n \geq 1$. This assumption is consistent with (\ref{GAMMA}).    
In the asymptotic limit $\tau_j \rightarrow \infty, 
\ \ j = 1,2,\cdots M$ with all $\tau_j - \tau_l, 
\ \ j,l = 1,2,\cdots,M$ fixed, it can be seen from 
(\ref{SEVENS}) that 
\begin{equation}
S(x,y;\tau,\tau^{\prime}) \sim S_1(x,y;\tau,\tau^{\prime}) + O({\rm e}^{-(\gamma_{(N/2)+1} - \gamma_{N/2})\tau}). 
\end{equation}
We can further derive the asymptotic relations $R_n(x;\tau) \sim O(1)$, 
$R^*_n(x;\tau) \sim O(1)$, 
$\Phi_n(x;\tau) \sim O({\rm e}^{- 2 \gamma_{n+1} \tau})$ and 
$\Phi^*_n(x;\tau) \sim O({\rm e}^{- 2 \gamma_{n+1} \tau})$ 
from (\ref{REVENS}) and (\ref{PEVENS}) so that 
\begin{equation}
I(x,y;\tau,\tau^{\prime}) \sim O({\rm e}^{- \gamma_{(N/2)+1} 
(\tau + \tau^{\prime})}), \ \ \ 
D(x,y;\tau,\tau^{\prime}) \sim O({\rm e}^{\gamma_{N/2} 
(\tau + \tau^{\prime})}). 
\end{equation}
Therefore in the asymptotic limit we obtain a quaternion determinant in 
which ${\tilde S}_{jl}^{mn}$ are replaced by 
\begin{equation}
\sigma_{jl}^{mn} = \left\{ \begin{array}{ll} 
S_1(\theta^m_j,\theta^n_l ;\tau_m,\tau_n) 
- g_{jl}^{mn}, \ \ \ m>n, \\ 
S_1(\theta^m_j,\theta^n_l ;\tau_m,\tau_n), 
\ \ \ m \leq n 
\end{array} \right.
\label{SIG}
\end{equation} 
and the other elements are set to zero. Then the 
quaternion determinant becomes an ordinary 
determinant 
\begin{eqnarray}
& & \rho(x^1_1,\cdots,x^1_{m_1}; \tau_1; x^2_1,\cdots,x^2_{m_2}; \tau_2; \cdots;x^M_1,\cdots,x^M_{m_M}; \tau_M) 
\nonumber \\
 & = & \left| \begin{array}{cccc} 
\sigma^{1 1}(m_1, m_1) &  \sigma^{1 2}(m_1, m_2) & \cdots &  \sigma^{1 M}(m_1, m_M) \\ 
\sigma^{2 1}(m_2, m_1) &  \sigma^{2 2}(m_2, m_2) & \cdots &  \sigma^{2 M}(m_2, m_M) \\  
\vdots & \vdots & \ddots & \vdots \\ 
\sigma^{M 1}(m_M, m_1) &  \sigma^{M 2}(m_M, m_2) & \cdots &  \sigma^{M M}(m_M, m_M) 
\end{array} \right|, \nonumber \\ 
\label{SIGDET}
\end{eqnarray}
where each block $\sigma^{\mu \nu}(m_{\mu}, m_{\nu})$ is obtained 
by removing the $m_{\mu} + 1, m_{\mu} + 2,  \cdots, N$-th rows 
and $m_{\nu} + 1, m_{\nu} + 2, \cdots, N$-th columns 
from $\sigma^{\mu \nu}$. This is the unitary symmetry version of 
a similar determinant formula \cite{EM,NFL} known for the Brownian dynamics 
within hermitian symmetry. 

\subsection{The Case $N$ Odd}
In the case with odd $N$, we introduce expansions ($n \geq 1$, 
$z={\rm e}^{i \theta}$) 
\begin{eqnarray}
\Omega_n(\theta;\tau) & = & {\rm e}^{- \gamma_n \tau} 
\sum_{j=-n+1}^n \alpha_{nj} z^j {\rm e}^{\gamma_j \tau}, \nonumber \\  
\Omega^*_n(\theta;\tau) & = & {\rm e}^{- \gamma_{-n} \tau} 
\sum_{j=-n+1}^n \alpha^*_{nj} z^{-j} {\rm e}^{\gamma_{-j} \tau} 
\end{eqnarray}
with $\alpha_{nn} = \alpha^*_{nn}=1$ and
\begin{equation}
\Omega_0(\theta;\tau) = {\rm e}^{- \gamma_0 \tau} 
\sum_{j=-(N-1)/2}^{(N-1)/2} \alpha_{0j} z^{j} {\rm e}^{\gamma_j \tau}. 
\end{equation}
As before it can be easily proven that
\begin{equation}
r_n(\tau)  = {\rm e}^{-\gamma_{-n} \tau} {\rm e}^{-\gamma_n \tau} 
r_n(0)
\end{equation}
and
\begin{equation}
s_n(\tau)  =  {\rm e}^{-\gamma_n \tau} s_n(0), \ \ \ 
s^*_n(\tau)  = {\rm e}^{-\gamma_{-n} \tau} s^*_n(0).
\end{equation}
Let us now define
\begin{eqnarray}
& & \Pi^*_n(\theta;\tau) = \int_{-\pi}^{\pi} 
F(\varphi,\theta;\tau,\tau) 
\Omega^*_n(\varphi;\tau) {\rm d}\varphi \nonumber \\ 
\displaystyle 
& - &  \left\{ \begin{array}{ll} 0, & 0 \leq n \leq (N-1)/2, \\ 
\displaystyle \frac{{\rm e}^{- \gamma_{-n} \tau}}{2 \pi} 
\left. \langle \Omega^*_n(\varphi;0),\Omega_0(\varphi^{\prime};0) 
\rangle \right|_{\tau=0} \Upsilon(\theta;\tau), & n \geq (N+1)/2,
\end{array} \right.  
\nonumber \\ 
& & \Pi_n(\theta;\tau) = \int_{-\pi}^{\pi} F(\varphi,\theta;\tau,\tau) 
\Omega_n(\varphi;\tau) {\rm d}\varphi \nonumber \\ 
\displaystyle 
& - &  \left\{ \begin{array}{ll} 0, & 0 \leq n \leq (N-1)/2, \\ 
\displaystyle \frac{{\rm e}^{- \gamma_n \tau}}{2 \pi} 
\left. \langle \Omega_n(\varphi;0),\Omega_0(\varphi^{\prime};0) 
\rangle \right|_{\tau=0} \Upsilon(\theta;\tau), & n \geq (N+1)/2,
\end{array} \right. \nonumber \\  
\end{eqnarray}
where
\begin{equation}
\Upsilon(\theta;\tau) = \sum_{l=0}^{\infty} z^{-l} 
{\rm e}^{- \gamma_l \tau} \lambda_{l0} 
+  \sum_{l=1}^{\infty} z^l 
{\rm e}^{- \gamma_{-l} \tau} \lambda^*_{l0}.
\end{equation} 
The inverse expansions ($n \geq 1$) 
\begin{eqnarray}
z^n {\rm e}^{\gamma_n \tau} & = &  \sum_{j=0}^{n_{\lambda}} \lambda_{nj} 
{\rm e}^{\gamma_j \tau} \Omega_j(\theta;\tau) + \sum_{j=1}^{n_{\mu}} 
\mu_{nj} {\rm e}^{\gamma_{-j} \tau} \Omega^*_j(\theta;\tau),
\nonumber \\ 
z^{-n} {\rm e}^{\gamma_{-n} \tau} & = &  \sum_{j=0}^{n_\lambda} 
\lambda^*_{nj} {\rm e}^{\gamma_{-j} \tau} \Omega^*_j(\theta;\tau) 
+ \sum_{j=1}^{n_{\mu}} \mu^*_{nj} 
{\rm e}^{\gamma_j \tau} \Omega_j(\theta;\tau) 
\label{ZODD1}
\end{eqnarray}
and
\begin{equation}
z^0 {\rm e}^{\gamma_0 \tau} =  \sum_{j=0}^{(N-1)/2} 
\lambda_{0j} {\rm e}^{\gamma_j \tau} \Omega_j(\theta;\tau) 
+ \sum_{j=1}^{(N-1)/2} \mu_{0j} {\rm e}^{\gamma_{-j} \tau} 
\Omega^*_j(\theta;\tau),
\label{ZODD2}
\end{equation} 
where
\begin{eqnarray}
n_{\lambda} & = & n_{\mu}+1= n, \ \ \ {\rm if} \ n > (N-1)/2, \nonumber \\ 
n_{\lambda} & = & n_{\mu} = (N-1)/2, \ \ \ {\rm if} \ n \leq (N-1)/2,
\end{eqnarray} 
lead to
\begin{eqnarray}
& & \Pi^*_n(\theta;\tau) \nonumber \\  
& = & \left\{ \begin{array}{ll} 
\displaystyle {\rm e}^{-\gamma_{-n} \tau} 
\frac{r_n(0)}{2 \pi}  
\left[ \sum_{l = 0}^{\infty} 
z^{-l} {\rm e}^{-\gamma_l \tau} \lambda_{ln}  
+ \sum_{l = 1}^{\infty} 
z^l {\rm e}^{-\gamma_{-l} \tau} \mu^*_{ln} \right], & 
1 \leq n \leq (N-1)/2, \\  
\displaystyle {\rm e}^{-\gamma_{-n} \tau} 
\frac{r_n(0)}{2 \pi}  
\left[ \sum_{l = n}^{\infty} 
z^{-l} {\rm e}^{-\gamma_l \tau} \lambda_{ln}  
+ \sum_{l = n+1}^{\infty} 
z^l {\rm e}^{-\gamma_{-l} \tau} \mu^*_{ln} \right], 
& n \geq (N+1)/2,
\end{array} \right. \nonumber 
\end{eqnarray}
\begin{eqnarray}
& & \Pi_n(\theta;\tau) \nonumber \\  
& = & \left\{ \begin{array}{ll} 
\displaystyle - {\rm e}^{-\gamma_n \tau} 
\frac{r_n(0)}{2 \pi}  
\left[ \sum_{l = 0}^{\infty} 
z^{-l} {\rm e}^{-\gamma_l \tau} \mu_{ln}  
+ \sum_{l = 1}^{\infty} 
z^l {\rm e}^{-\gamma_{-l} \tau} \lambda^*_{ln} \right], & 
1 \leq n \leq (N-1)/2, \\  
\displaystyle - {\rm e}^{-\gamma_n \tau} 
\frac{r_n(0)}{2 \pi}  
\left[ \sum_{l = n+1}^{\infty} 
z^{-l} {\rm e}^{-\gamma_l \tau} \mu_{ln}  
+ \sum_{l = n}^{\infty} 
z^l {\rm e}^{-\gamma_{-l} \tau} \lambda^*_{ln} \right], & 
n \geq (N+1)/2. \end{array} \right. \nonumber \\ 
\label{PHIODD} 
\end{eqnarray}
Using the formula (\ref{PHIODD}) for $\Pi^*_n(\theta;\tau)$ 
and $\Pi_n(\theta;\tau)$, we can readily derive
\begin{equation}
f(\theta;\tau) = \sum_{n=1}^{\infty} \frac{1}{r_n(\tau)} 
[\Pi^*_n(\theta;\tau) s_n(\tau)  
- \Pi_n(\theta;\tau) s^*_n(\tau)] \nonumber \\ 
+ \frac{1}{2 \pi} s_0(0) \Upsilon(\theta;\tau).
\label{FODD}
\end{equation}   
Substituting (\ref{PHIODD}) and ({\ref{FODD}) into (\ref{SODD}) 
yields
\begin{equation}
S(\theta,\theta^{\prime};\tau,\tau^{\prime}) =  
S_1(\theta,\theta^{\prime};\tau,\tau^{\prime}) 
+ S_2(\theta,\theta^{\prime};\tau,\tau^{\prime}),
\label{SODDS}
\end{equation}
where ($z = {\rm e}^{i \theta}, \ z^{\prime} = 
{\rm e}^{i \theta^{\prime}}$)
\begin{equation}
S_1(\theta,\theta^{\prime};\tau,\tau^{\prime}) = 
\frac{1}{2 \pi} \sum_{n=-(N-1)/2}^{(N-1)/2} 
\left( \frac{z^{\prime}}{z} \right)^n 
{\rm e}^{-\gamma_n (\tau - \tau^{\prime})}
\label{S1ODD}
\end{equation}
and
\begin{eqnarray}
& & S_2(\theta,\theta^{\prime};\tau,\tau^{\prime}) 
= \frac{1}{2 \pi}  
\sum_{k=0}^{(N-1)/2} 
\sum_{\nu=(N+1)/2}^{\infty} \nonumber \\ 
& \times & \left[ \left(z^{-\nu} {\rm e}^{-\gamma_{\nu} \tau} 
\lambda_{\nu k} + z^{\nu} {\rm e}^{-\gamma_{-\nu} \tau} 
\mu^*_{\nu k}) {\rm e}^{\gamma_k \tau^{\prime}} 
\Omega_k(\theta^{\prime};\tau^{\prime}\right) \right. \nonumber \\ 
& +  & \left. \left(z^{-\nu} {\rm e}^{-\gamma_{\nu} \tau} 
\mu_{\nu k} + z^{\nu} {\rm e}^{-\gamma_{-\nu} \tau} 
\lambda^*_{\nu k}\right) {\rm e}^{\gamma_{-k} \tau^{\prime}} 
\Omega^*_k(\theta^{\prime};\tau^{\prime}) \right] \nonumber \\ 
& + & \frac{\Pi_0(\theta;\tau)}{s_0(\tau)} 
\sum_{k=1}^{(N-1)/2} \frac{1}{r_k(\tau^{\prime})} \left[ 
\Omega^*_k(\theta^{\prime};\tau^{\prime}) s_k(\tau^{\prime}) 
- \Omega_k(\theta^{\prime};\tau^{\prime}) s^*_k(\tau^{\prime}) 
\right] \nonumber \\  
& + & \frac{\Omega_0(\theta^{\prime};\tau^{\prime})}{s_0(\tau^{\prime})} 
\sum_{k=(N+1)/2}^{\infty} \frac{1}{r_k(\tau)} \left[ 
\Pi^*_k(\theta;\tau) s_k(\tau) 
- \Pi_k(\theta;\tau) s^*_k(\tau) 
\right], \nonumber \\  
\end{eqnarray} 
where we set $\mu_{\nu \ 0}= \mu^*_{\nu \ 0}=0$. 
Finally we put (\ref{ZODD1}) and (\ref{ZODD2}) into (\ref{FTHETA}) 
to find 
\begin{eqnarray}
& & I(\theta,\theta^{\prime};\tau,\tau^{\prime}) \nonumber \\ 
& = & \sum_{k=(N+1)/2}^{\infty} \frac{1}{r_k(\tau^{\prime})} 
\left[ {\rm e}^{-\gamma_{-k} (\tau^{\prime} - \tau)} 
\Pi^*_k(\theta;\tau) \Pi_k(\theta^{\prime};\tau^{\prime})  
-  {\rm e}^{-\gamma_k (\tau^{\prime}- \tau)} \Pi_k(\theta;\tau) 
\Pi^*_k(\theta^{\prime};
\tau^{\prime}) 
\right] \nonumber \\  
& + & \frac{\Pi_0(\theta;\tau)}{s_0(\tau)} 
\sum_{k=(N+1)/2}^{\infty} \frac{1}{r_k(\tau^{\prime})} 
\left[  
\Pi^*_k(\theta^{\prime};\tau^{\prime})  
s_k(\tau^{\prime}) - 
\Pi_k(\theta^{\prime};\tau^{\prime}) 
s^*_k(\tau^{\prime})
\right] \nonumber \\  
& - & \frac{\Pi_0(\theta^{\prime};
\tau^{\prime})}{s_0(\tau^{\prime})} 
\sum_{k=(N+1)/2}^{\infty} \frac{1}{r_k(\tau)} 
\left[ \Pi^*_k(\theta;\tau)  
s_k(\tau) - \Pi_k(\theta;\tau) 
s^*_k(\tau)
\right]. \nonumber \\ 
\end{eqnarray}
Assuming that $\gamma_n=\gamma_{-n}$ and $\gamma_{n+1} > \gamma_n$ 
for $n \geq 0$, we can now take the asymptotic limit 
$\tau_j \rightarrow \infty, \ \ j = 1,2,\cdots M$ 
with all $\tau_j - \tau_l, \ \ j,l = 1,2,\cdots,M$ 
fixed. Let us first note that $\Omega_n(x;\tau) \sim O(1)$, 
$\Omega^*_n(x;\tau) \sim O(1)$, $\Pi_n(x;\tau) \sim 
O({\rm e}^{- 2 \gamma_n \tau})$ and 
$\Pi^*_n(x;\tau) \sim O({\rm e}^{- 2 \gamma_n \tau})$ for $n \geq 1$. 
We can further obtain an estimation 
\begin{equation}
\Omega_0(\theta;\tau) \sim O({\rm e}^{-(\gamma_0 - 
\gamma_{(N-1)/2}) \tau}), \ \ \   
\Pi_0(\theta;\tau) \sim O({\rm e}^{- (\gamma_0 + 
\gamma_{(N+1)/2}) \tau}). 
\end{equation}
Then it is straightforward to find
\begin{equation}
S(x,y;\tau,\tau^{\prime}) \sim 
S_1(x,y;\tau,\tau^{\prime}) + 
O({\rm e}^{-(\gamma_{(N+1)/2} - \gamma_{(N-1)/2})\tau}), 
\end{equation}
\begin{equation}
I(x,y;\tau,\tau^{\prime}) \sim O({\rm e}^{- \gamma_{(N+1)/2} 
(\tau + \tau^{\prime})}), \ \ \ 
D(x,y;\tau,\tau^{\prime}) \sim O({\rm e}^{\gamma_{(N-1)/2} 
(\tau + \tau^{\prime})}). 
\end{equation}
Therefore $I^{mn}_{jl}$ and $D^{mn}_{jl}$ can be set 
to zero in the asymptotic limit and, as in the case $N$ even, 
a determinant expression (\ref{SIGDET}) results. Here 
the matrix elements $\sigma^{mn}_{jl}$ are again defined by (\ref{SIG}), 
although the definitions of $S_1(\theta,\theta^{\prime};\tau,\tau^{\prime})$ 
and $g(\theta,\varphi;\tau)$ are different from those 
in the case $N$ even.  

\section{The Case $U(\theta)=1$}
\setcounter{equation}{0}
\renewcommand{\theequation}{4.\arabic{equation}}
In this section we deal with the simplest case 
$U(\theta)=1$ and explicitly give the formulas for 
skew orthogonal polynomials. Though this case was 
already treated by Pandey and Shukla, our result is 
more general than theirs because they evaluated only 
the equal time correlations.  

\subsection{The Case $N$ Even}
\par
\noindent
\medskip
(1) Symmetric unitary initial condition
\par
\noindent
\medskip
For the bracket defined in (\ref{SKEW}) corresponding 
to the first of the three initial conditions (\ref{INITIAL}), 
we can easily derive ($z = {\rm e}^{i \theta}$)
\begin{equation}
\left. \langle z^{m+(1/2)},z^{n+(1/2)} \rangle \right|_{\tau=0} 
= \frac{4 \pi i}{n + (1/2)} 
\delta_{m + n + 1 \ 0},
\end{equation}
which yields
\begin{equation}
R_n(\theta;0) = z^n, \ \ R^*_n(\theta;0) = z^{-n}, \ \ r_n(0) = 
\frac{4 \pi i}{n + (1/2)}. 
\end{equation}
\par
\noindent
\medskip
(2) Self-dual quaternion unitary initial condition
\par
\noindent
\medskip
Let us next consider the second of the initial 
conditions (\ref{INITIAL}). Noting 
\begin{equation}
\left. \langle z^{m+(1/2)},z^{n+(1/2)} \rangle \right|_{\tau=0} 
= 4 \pi i \left( n + \frac{1}{2} \right) \delta_{m + n + 1 \ 0},
\end{equation}
we obtain
\begin{equation}
R_n(\theta;0) = z^n, \ \ R^*_n(\theta;0) = z^{-n}, \ \ r_n(0) = 
4 \pi i \left( n + \frac{1}{2} \right).
\end{equation}
\par
\noindent
\medskip
(3) An analogue of antisymmetric hermitian initial condition 
\par
\noindent
\medskip
The third of the initial conditions (\ref{INITIAL}) yields 
\begin{equation}
\left. \langle z^{m+(1/2)},z^{n+(1/2)} \rangle \right|_{\tau=0} 
= \left\{ \begin{array}{ll} 
\pi i, & {\rm if} \ n-m \ {\rm is} \ {\rm odd} \ 
{\rm and} \ {\rm positive}, \\  
0, & {\rm if} \ n-m \ {\rm is} \ {\rm even}, \\ 
- \pi i, & {\rm if} \ n-m \ {\rm is} \ {\rm odd} \ 
{\rm and} \ {\rm negative}, \end{array} \right.
\end{equation}
which implies ($n \geq 1$)
\begin{equation}
R_n(\theta;0) = z^n + z^{-n}, \ \ R^*_n(\theta;0) = z^{-n} + z^n, \ \ 
r_n(0) = 2 \pi i
\end{equation} 
and
\begin{equation}
R_0(\theta;0) = 1, \ \ r_0(0) = \pi i.
\end{equation} 
\subsection{The Case $N$ Odd}
\par
\noindent
\medskip
(1) Symmetric unitary initial condition
\par
\noindent
\medskip
In $N$ odd case of the symmetric unitary initial condition, 
we obtain 
\begin{equation}
\left. \langle z^m,z^n \rangle \right|_{\tau=0} 
= \frac{4 \pi i}{n} 
\delta_{m + n \ 0}, \ \ \ m,n \neq 0
\end{equation}
and
\begin{equation}
\left. \langle z^n,1 \rangle \right|_{\tau=0} 
= \frac{4 \pi i}{n} (-1)^n. 
\end{equation}
Then, putting $a_n=a^*_n=0$, we can derive 
($n \geq 1$)
\begin{equation}
\Omega_n(\theta;0) = z^n, \ \ \Omega^*_n(\theta;0) = z^{-n}, \ \ r_n(0) = 
\frac{4 \pi i}{n} 
\end{equation}
and
\begin{equation}
\Omega_0(\theta;0) = \sum_{j=-(N-1)/2}^{(N-1)/2} (-z)^j.
\end{equation}
\par
\noindent
\medskip
(2) An analogue of antisymmetric hermitian initial condition 
\par
\noindent
\medskip
An analogue of the antisymmetric hermitian initial condition 
is handled by considering the inner products 
\begin{equation}
\left. \langle z^m,z^n \rangle \right|_{\tau=0} 
= \left\{ \begin{array}{ll} 
\pi i, & {\rm if} \ n-m \ {\rm is} \ {\rm odd} \ 
{\rm and} \ {\rm positive}, \\  
0,  & {\rm if} \ n-m \ {\rm is} \ {\rm even}, \\ 
- \pi i, & {\rm if} \ n-m \ {\rm is} \ {\rm odd} \ 
{\rm and} \ {\rm negative}. \end{array} \right.
\end{equation}
We fix the constants as $a_n = a^*_n = (-1)^{n+1}$ and find 
($n \geq 1$)
\begin{equation}
\Omega_n(\theta;0) = z^n + z^{-n+1}, 
\ \ \Omega^*_n(\theta;0) = z^{-n} + z^{n-1}, \ \ 
r_n(0) = 2 \pi i
\end{equation} 
and
\begin{equation}
\Omega_0(\theta;0) = (-1)^{(N-1)/2} \frac{1}{2} (z^{(N-1)/2} + z^{-(N-1)/2}). 
\end{equation} 

\section{Conclusion}
\setcounter{equation}{0}
\renewcommand{\theequation}{5.\arabic{equation}}
In this paper we have derived quaternion determinant 
expressions for the dynamical multilevel correlation functions 
for Dyson's Brownian motion of eigenparameters (energy 
levels) toward unitary symmetry. As the initial conditions, 
we assume eigenparameter distributions of symmetric unitary, 
self-dual quaternion unitary and an analogue of the antisymmetric 
hermitian random matrices. Any of the quaternion elements is 
represented in terms of four functions $D(\theta,\theta^{\prime};
\tau,\tau^{\prime})$, $I(\theta,\theta^{\prime};\tau,\tau^{\prime})$, 
$S(\theta,\theta^{\prime};\tau,\tau^{\prime})$ and 
$g(\theta,\theta^{\prime};\tau - \tau^{\prime})$ appearing 
in the two level dynamical correlation functions. Therefore 
analyzing the two level dynamical correlations is enough to 
understand the behavior of all the multilevel correlations. 
The two level dynamical correlations are investigated 
in Refs.\cite{SPO,PJF2,FNJSP} and some asymptotic results in the limit 
$N \rightarrow \infty$ are already known.   

\bibliographystyle{plain}

\end{document}